\documentclass[12pt,a4paper]{article}
\usepackage{color}
\usepackage{amsmath,amssymb,amsthm}
\newtheorem{theorem}{Theorem}
\newtheorem{lemma}{Lemma}
\newtheorem{definition}{Definition}
\newtheorem{proposition}{Proposition}

\usepackage{natbib} 
\usepackage{algorithm,algorithmic}
\usepackage{amsmath,amssymb}
\usepackage{bm}
 \bmdefine\g{g} \bmdefine\K{K}\bmdefine\X{X}
\bmdefine\X{X} \bmdefine\D{D} \bmdefine\K{K} \bmdefine\Z{Z}
\bmdefine\x{x} \bmdefine\z{z} \bmdefine\y{y} \bmdefine\Y{Y}
\bmdefine\bfalpha{\alpha} \bmdefine\bfmu{\mu} \bmdefine\M{M}
\bmdefine\Q{Q} \bmdefine\P{P} \bmdefine\w{w} \bmdefine\W{W}
\bmdefine\p{p} \bmdefine\T{T} \bmdefine\t{t} \bmdefine\r{r}
\bmdefine\B{B} \bmdefine\I{I} \bmdefine\u{u} \bmdefine\p{p}
\bmdefine\Sig{\Sigma} \bmdefine\E{E} \bmdefine\F{F} \bmdefine\S{S}
\bmdefine\s{s} \bmdefine\w{w} \bmdefine\b{b} \bmdefine\W{W}
\bmdefine\w{w} \bmdefine\V{V} \bmdefine\v{v} \bmdefine\q{q}
\bmdefine\R{R} \bmdefine\A{A} \bmdefine\bflambda{\lambda}
\bmdefine\C{C} \bmdefine\U{U} \bmdefine\L{L} \bmdefine\d{d}
\bmdefine\c{C}
\newcommand{\Hc}{{\mathcal{H}}}
\newcommand{\Kc}{{\mathcal{K}}}
\newcommand{\Lc}{{\mathcal{L}}}
\newcommand{\Pc}{{\mathcal{P}}}
\newcommand{\Xc}{{\mathcal{X}}}

\newcommand{\Bc}{{\mathcal{B}}}
\newcommand{\Yc}{{\mathcal{Y}}}
\newcommand{\mbe}{\mathbb{E}}
\newcommand{\mbr}{\mathbb{R}}

\newcommand{\norm}[1]{\left\|#1\right\|}
\newcommand{\brac}[1]{\left[#1\right]}
\newcommand{\e}[1]{\mbe\brac{#1}}
\newcommand{\inner}[1]{\left\langle#1\right\rangle}
\newcommand{\eps}{\varepsilon}

\newcommand{\paren}[1]{\left(#1\right)}
\newcommand{\abs}[1]{\left|#1\right|}

\begin{document}



\title{Kernel Partial Least Squares \\is Universally Consistent}
\author{Gilles Blanchard\\{\small{Weierstrass Institute for}} \\{\small{Applied Analysis and Stochastics}}\\
 \small{Mohrenstr. 39}\\
 \small{10117 Berlin, Germany}
 \and Nicole Kr\"amer\\\small{Berlin Institute of Technology}\\\small{Machine Learning Group}\\\small{Franklinstr. 28/29}\\\small{10587 Berlin, Germany} }
\maketitle

\begin{abstract}
We prove the statistical consistency of kernel Partial Least Squares Regression applied to
a bounded regression learning problem on a reproducing
kernel Hilbert space.
Partial Least Squares  stands out of well-known classical approaches as e.g. Ridge Regression or Principal Components Regression, as it is not defined as the solution of a global cost minimization procedure over a fixed model nor is it a linear estimator. Instead, approximate solutions are constructed by projections onto a nested set of data-dependent subspaces. To prove consistency, we exploit the known fact that Partial Least Squares is equivalent to the conjugate gradient algorithm in combination with early stopping. The choice of the stopping rule (number of iterations) is a crucial point. We study two empirical stopping rules. The first one monitors the estimation error in each iteration step of Partial Least Squares, and the second one estimates the empirical complexity in terms of a condition number. Both stopping rules lead to universally consistent estimators provided the kernel is universal.
\end{abstract}

\section{INTRODUCTION}
Partial Least Squares (PLS) \citep{Wold7501,Wold8401} is a
supervised  dimensionality reduction technique. It iteratively
constructs an orthogonal set of $m$ latent components from the predictor variables  which have maximal
covariance with the response variable. This low-dimensional representation of the data  is then used for prediction by fitting a linear regression model to the response and the latent components. The number $m$ of latent components acts as a regularizer. In contrast to Principal Components Regression, the latent components are response-dependent. 
In combination with the kernel trick \citep{SchSmoMue98}, kernel PLS
performs nonlinear dimensionality reduction and regression \citep{Rosipal0101}.

While PLS has proven to be successful in a wide range of applications, theoretical studies of PLS -- such as its consistency -- are less widespread. This is perhaps due to the fact that in contrast to many standard methods (as e.g.~Ridge Regression or Principal Components Regression), PLS is not defined as the solution of a global cost function nor is it a linear estimator in the sense that the fitted values depend linearly on the response variable. Instead, PLS minimizes the least squares criterion on a nested subset of data-dependent subspaces (i.e., the subspaces defined by the latent components). Therefore, results obtained for linear estimators are not straightforward to extend to PLS. Recent work \citep{Naik0001,Chun0901} study the model consistency of PLS in the linear case. Their results assume that the target function depends on a finite known number $\ell$ of orthogonal latent components and that PLS is run at least for $\ell$ steps (without early stopping).
In this configuration,  \citet{Chun0901} obtain inconsistency results  in scenarios where the dimensionality can grow with the number of data. This underscores that the choice of the regularization (or early stopping) term $m$ is important and that it has to be selected in a data-dependent manner.

Here, we prove the universal prediction consistency of kernel PLS in the infinite dimensional
case. In particular, we define suitable data-dependent stopping criteria for the number of PLS components to
ensure consistency. For the derivation of our results, we capitalize on the close connection of PLS and the conjugate gradient algorithm \citep{Hestenes5201} for the solution of linear equations: The PLS solution with $m$ latent components is equivalent to the conjugate algorithm applied to the set of normal equations in combination with early stopping after $m$ iterations. We use this equivalence to define the population version of kernel PLS. We then proceed in three steps: (i) We show that population kernel PLS converges to the true regression function. (ii) We bound the difference between empirical and  population PLS, which is low as long as the number of iterations does not grow too fast. We ensure this via two different stopping criteria. The first one monitors the error in each iteration stop of PLS, and the second one estimates the empirical complexity in terms of a
condition number. (iii) Combining the results from the two previous steps, our stopping rules lead to universally consistent estimators provided the kernel is universal. We emphasize that either stopping rule does not depend on any prior knowledge of the target function and only depends on observable quantities.

\section{BACKGROUND}
We study a regression problem  based on a joint probability distribution $P(X,Y)$ on $\Xc \times \Yc$. The task is to estimate the true regression function
\begin{eqnarray}
\label{eq:reg}
\bar{f}(x) &=& \e{Y|X=x}
\end{eqnarray}
based on a finite  number $n$ of  observations $\left(x_1,y_1\right),\ldots,\left(x_n,y_n\right)\in \mathcal{X} \times \mathcal{Y}$. As a general convention, population quantities defined from the perfect knowledge of the distribution $P$ will be denoted with a bar, empirical quantities without.
We assume that $\bar{f}$ belongs
to the space of $P_X$-square-integrable functions $\Lc_2(P_X)$\,, where $P_X$ denotes the $X$-marginal distribution.
The vector $\y \in \mathbb{R}^n$ represents the $n$ centered response observations $y_1,\ldots,y_n$.

As we consider kernel techniques to estimate the true regression function \eqref{eq:reg}, we map the data
to a reproducing kernel Hilbert space $\Hc_k$ with bounded kernel
$k$, via the canonical kernel mapping
\begin{displaymath}
\phi(x): \Xc \rightarrow \Hc_k, \quad
x \mapsto \phi(x) = k(x,\cdot)\,.
\end{displaymath}
In the remainder, we make the following assumptions:

{\bf (B)} boundedness of the data and kernel: $Y\in [-1,1]$ almost
surely and $\sup_{x\in \Xc} k(x,x) \leq 1$\,.

{\bf (U)} universality of the kernel: for any distribution $P_X$ on
$\Xc$\,, $\Hc_k$ is dense in $\Lc_2(P_X)$\,.

\subsection{LEARNING AS AN INVERSE PROBLEM}
We very briefly review the interpretation of Kernel based regression as a statistical inverse problem, as introduced in \cite{Vito0601}. Let us denote the inclusion operator of the kernel space into $\mathcal{L}_2(P_X)$ by
\begin{eqnarray*}
\bar{T}:\Hc_k\hookrightarrow \Lc_2(P_{X})\,&&g\mapsto g\,.
\end{eqnarray*}
This operator maps a function to itself, but between two Hilbert spaces which differ with respect to their geometry  -- the inner product of $\Hc_k$ being defined by the kernel function $k$, while the inner product of $\mathcal{L}_2(P_X)$ depends on the data generating distribution. The adjoint operator of $\bar{T}$ is defined as usual as the unique operator satisfying the condition
\begin{eqnarray*}
\inner{f,\bar{T}g}_{\mathcal{L}_2(P_X)} &=& \inner{\bar{T}^* f,g}_{\Hc_k}
\end{eqnarray*}
for all $f \in \mathcal{L}_2(P_X), g\in \Hc_k$.
It can be checked from this definition and the reproducing property of the kernel that $\bar{T}^*$ coincides with the {\em kernel integral operator} from $\mathcal{L}_2(P_X)$ to $\Hc_k$\,,
\begin{eqnarray*}
\bar{T}^*g&=&\int k(.,x') g(x') dP(x') = \e{k(X,\cdot) g(X)}\,.
\end{eqnarray*}
Finally, the operator $\bar{S}=\bar{T}^* \bar{T}:\Hc_k\rightarrow \Hc_k$ is the {\em covariance operator} for the random variable $\phi(X)$:
\[
Sg = \e{k(X,\cdot) g(X)} = \e{\phi(X) \inner{\phi(X),g}}\,.
\]
Learning in the kernel space $\Hc_k$
can be cast (formally) as the inverse problem $\bar{T}\bar{g}=\bar{f}$, which yields (after right multiplication by $\bar{T}^*$) the so-called normal equation
\begin{eqnarray}\label{eq:normal}
\bar{S} \bar{g}&=& \bar{T}^* \bar{f}\,.
\end{eqnarray}
The above equation has a solution if and only if $\bar{f}$ can be represented as a function of $\Hc_k$, i.e.~$\bar{f} \in \bar{T}\Hc_k$\,; however, even if
$\bar{f} \not\in \bar{T}\Hc_k$, the above formal equation can be used as a motivation to use regularization algorithms coming from the inverse problems literature in order to find an {\em approximation} $\bar{g}$ of $\bar{f}$ belonging to the space $\Hc_k$.
In a learning problem, neither the left-hand nor the right-hand side of \eqref{eq:normal} is known, and we only observe empirical quantities, which can be interpreted as ''perturbed'' versions of the population equations wherein $P_X$ is replaced by its empirical counterpart $P_{X,n}$ and $\bar{f}$ by $\y$\,. Note that the space $\mathcal{L}_2(P_{X,n})$ is isometric to $\mathbb{R}^n$ with the usual Euclidean product, wherein a function $g\in\mathcal{L}_2(P_{X,n})$ is mapped to the $n$-vector $(g(x_1),\ldots,g(x_n))$.
The empirical  integral operator
$T^* : \Lc_2(P_{X,n})  \rightarrow  \Hc_k$ is then given by
\begin{eqnarray}
\label{eq:kernel_emp}
T^* g &=& \frac{1}{n} \sum_{i=1}^n g(x_i) k(x_i,\cdot)\,.
\end{eqnarray}
The empirical covariance operator $S=T ^*T$ is defined similarly, but on the input space $\Hc_k$\,. Note that if the Hilbert space $\Hc_k$ is finite-dimensional, the operator $S$ corresponds to left multiplication with the empirical covariance matrix. The perturbed, empirical version of the normal equation \eqref{eq:normal} is then defined as
\begin{equation}
\label{eq:normal_emp}
S g = T^*  \y\,.
\end{equation}
Again, if $\Hc_k$ is finite-dimensional, the right-hand side corresponds to the covariance between predictors and response. In general,  equation \eqref{eq:normal_emp} is ill-posed, and regularization techniques are needed. Popular examples are Kernel Ridge Regression (which corresponds to Tikonov regularization in inverse problems) or $\ell_2$-Boosting (corresponding to Landweber iterations).

\subsection{PLS AND CONJUGATE GRADIENTS}\label{subsec:cg} %

PLS is generally described as a greedy iterative method that produces a sequence of latent components on which the data is projected.
In contrast with PCA components, which maximize the variance, in PLS, the components are defined to have maximal covariance with the response $\y$.  In particular, the latent components depend on the response. For prediction, the response $\y$ is projected on these latent components.

 It is however a known fact that the output of the $m$-th step of PLS is actually equivalent to a conjugate gradient (CG) algorithm applied to the normal equation \eqref{eq:normal_emp}, stopped early at step $m$ (see e.g. \citealp{Phatak0201} for a detailed overview). This has been established for traditional PLS, i.e.~$\mathcal{X} = \mathbb{R}^d$ and the linear kernel is used, but for the kernel PLS (KPLS) algorithm introduced by \cite{Rosipal0101} the exact same analysis is valid as well. Here, for reasons of clarity with the remainder of our analysis we therefore directly present KPLS as a CG method.

 For the self-adjoint operator $S$  and for $T^*\y$, we define the associated Krylov space of order $m$ as
\begin{eqnarray*}
\Kc_m\left(T^*\y, S\right)= \text{span}\left\{T^*\y,S T^*\y,\ldots,S^{m-1} T^*\y\right\} \subset \Hc_k\,.
\end{eqnarray*}
In other words, $\Kc_m\left(T^* \y, S\right)$ is the linear subspace of $\Hc_k$ of all elements of the form $q(S)T^* \y$\,, where $q$ ranges over the real polynomials of degree $m-1$\,.
The $m$-th step of the CG method as applied to the normal equation \eqref{eq:normal_emp} is simply defined (see e.g. \citealp{Engl}, chap. 7) as the element $g_m \in \Hc_k$ that minimizes the least squares criterion over the Krylov space, i.e.
\begin{eqnarray}
\label{eq:cg_crit}
g_m&=& \text{arg} \min_{g \in \mathcal{K}_m(T^*\y,S)} \left\| \y- Tg \right\|^2\,.
\end{eqnarray}
Here, we recall that for any function $g\in \Hc_k$, the mapping $Tg$ of $g$ into $\mathcal{L}_2(P_{X,n})$ can be equivalently represented as the $n$-vector $(g(x_1),\ldots,g(x_n))$\,. Observe that since the Krylov space depends itself on the data (and in particular on the response variable $\y$), \eqref{eq:cg_crit} is not a linear estimator.


An extremely important property of CG is that the above optimization problem can be exactly computed by a simple iterative algorithm which only requires to use forward applications of the operator $S$\,, sums of elements in $\Hc_k$ and scalar multiplications and divisions (algorithm \ref{algo:cg1}).

\begin{algorithm}
\caption{Empirical KPLS (in  $\Hc_k$)}
\begin{algorithmic}
\STATE{Initialize:
$g_0 =0;  u_0 = T ^*\y ; d_0 = u_0 $}
\FOR{$m=0,\ldots,(m_{\max}-1)$}
\STATE{$\alpha_m = \norm{u_m}^2/\inner{d_m,Sd_m}$}
\STATE{$g_{m+1} = g_m + \alpha_m d_m$ (update)}
\STATE{$u_{m+1} = u_m - \alpha_m S d_m$ (residuals)}
\STATE{$\beta_m = \norm{u_{m+1}}^2/\norm{u_{m}}^2$}
\STATE{$d_{m+1} = u_{m+1} + \beta_m d_m$ (new basis vector)}
\ENDFOR
\STATE{{\bf Return:} approximate solution $g_{m_{\max}}$}
\end{algorithmic}
\label{algo:cg1}
\end{algorithm}
In fact, the CG algorithm iteratively constructs a basis $d_0,\ldots,d_{m-1}$  of the Krylov space $\Kc_m\left(T^*\y, S\right)$.
The sequence of $g_m \in \Hc_k$ is constructed in such a way that the residuals $u_m = T^*\y - S g_m$ are pairwise orthogonal in $\Hc_k$, i.e.~$\inner{u_j, u_k} =0$ for $j\not =k$, while the constructed basis is $S$-orthogonal (or equivalently, uncorrelated), i.e.~$\inner{d_j, S d_k} =0$ for $j\not =k$.

Note that the above algorithm is written entirely in $\Hc_k$, this form being convenient for the theoretical analysis to come. In practice, since all involved elements belong to $\mathrm{span}\{ (K(x_i,.))_{1\leq i \leq n} \}$, a weighted kernel expansion is used to represent these elements, and  corresponding weight update equations using the kernel matrix can be derived (see \citealp{Rosipal0101}).

\section{POPULATION VERSION OF KPLS}

Using the CG interpretation of KPLS, we can define its population version as follows:

\begin{definition}Denote by $\bar{g}_m\in \Hc_k$ the output of algorithm \ref{algo:cg1} after $m$ iterations, if we replace the empirical operator $S$ and the vector $T^* \y$  by their population versions $\bar{S}$ and $\bar{T}^* \bar{f}$, respectively. We define population KPLS with $m$ components as $\bar{f}_m = \bar{T} \bar{g}_m \in \mathcal{L}_2(P_X)$\,.
\end{definition}
We emphasize again that $\bar{g}_m\in \Hc_k$ and $\bar{f}_m\in \mathcal{L}_2(P_X)$ are identical as functions from $\Xc$ to $\mathbb{R}$, but seen as elements of Hilbert spaces with a different geometry (norm).
The first step in our consistency proof is to show that population KPLS $\bar{f}_m$ converges to $\bar{f}$ (with respect to the $\mathcal{L}_2(P_X)$ norm) if $m$ tends to $\infty$. Note that even if
$\bar{f} \not\in \bar{T}\Hc_k$,
we can still show that $\bar{T}\bar{g}_m$ converges to the projection of $\bar{f}$ onto the closure of $\Hc_k$ in
 $\Lc_2(P)$\,. If the kernel is universal {\bf(U)}, this projection is $\bar{f}$ itself and
 this implies asymptotic consistency of the population version.
We  will assume for
 simplicity that

{\bf (I)} the true regression function $\bar{f}$ has infinitely many components
in its decomposition over the eigenfunctions of $\bar{S}$\,,

which implies that the population version of the algorithm can
theoretically be run indefinitely without exiting. If this condition
is not satisfied the population algorithm stops after a finite number of steps
$\kappa$\,, at which points it holds that $\bar{f}_\kappa = \bar{f}$\,
so that the rest of our analysis also holds in that case with only minor modifications.
\begin{proposition}
\label{lem:opt2}
The kernel operator of $k$ is defined as $\bar{K}= \bar{T} \bar{T}^* :\Lc_2(P_X) \rightarrow  \Lc_2(P_X)$.
We denote by $\Pc$ the orthogonal projection onto the closure of the range of  $\bar{K}$ in $\Lc_{2}(P)$. Then, recalling $\bar{f}_m = \bar{T} \bar{g}_m$ where
$\bar{g}_m$ is the output of the $m$-th iteration of
the conjugate gradient algorithm applied to the population normal equation
\eqref{eq:normal}, it holds that $\bar{f}_m = \bar{K}q_m(\bar{K})\bar{f}$\,, where $q_m$ is a polynomial of degree $\leq m-1 $ fulfilling
\begin{eqnarray*}
q_m&=& \mathrm{arg}\min_{\deg q \leq m-1} \| \Pc \bar{f} - \bar{K} q(\bar{K}) \bar{f}\|^2_{\mathcal{L}_2(P_X)}\,.
\end{eqnarray*}
\end{proposition}
\begin{proof}
The minimization property \eqref{eq:cg_crit} when written in the population case yields
\begin{eqnarray*}
q_m&=& \text{arg}\min_{\deg q \leq m-1} \|\bar{f} - \bar{T} q(\bar{S}) \bar{T}^* \bar{f}\|^2\,.
\end{eqnarray*}
Furthermore,  for all polynomials $q$
\begin{eqnarray*}
\norm{\bar{f} - \bar{T} q(\bar{S}) \bar{T}^*\bar{f}}^2
& = &\norm{\bar{f} - \bar{K} q(\bar{K}) \bar{f}}^2\\
&= &\|\Pc \left(f-  \bar{K} q(\bar{K})\bar{f}\right)\|^2  + \|  \left(I-\Pc\right) \left(\bar{f}- \bar{K} q(\bar{K})\bar{f} \right) \|^2\\
&= &\| \Pc \bar{f} - \bar{K} q(\bar{K})\bar{f}\|^2 + \|\left(I-\Pc\right) \bar{f}\|^2\,.
\end{eqnarray*}
As the second term above does not depend on the polynomial $q$,
this yields the announced result.
\end{proof}

This leads to the following convergence result.

\begin{theorem}
\label{th:popcase}
Let us denote by $\widetilde f_m$ the projection of $\bar{f}$ onto the first $m$ principal components of the operator $\bar{K}$. We have
\begin{eqnarray*}
\| \Pc \bar{f} - \bar{f}_m\|_{\mathcal{L}_2(P_X)}&\leq \|\Pc \bar{f} - \widetilde{f}_m\|_{\mathcal{L}_2(P_X)}\,.
\end{eqnarray*}
In particular,
\begin{eqnarray*}
\bar{f}_m &\stackrel{m\rightarrow \infty}{\longrightarrow}& \Pc \bar{f} \quad \text{ in } \Lc_2(P_X)\,.
\end{eqnarray*}
\end{theorem}
This theorem is an extension of the finite-dimensional results by \cite{Phatak0201}.

\begin{proof}
 We construct a sequence of polynomials $\widetilde q_m $ of degree $\leq m-1$ such that
\begin{eqnarray*}
\|\Pc \bar{f} - \bar{K} \widetilde q_m(\bar{K}) \bar{f}\|&\leq &\|\Pc \bar{f} - \widetilde{f}_m\|
\end{eqnarray*}
 and then exploit the minimization property of Proposition \ref{lem:opt2}. Let us consider the first $m$ eigenvalues $\lambda_1,\ldots,\lambda_m$ of the operator $\bar{K}$ with corresponding eigenfunctions $\phi_1,\ldots,\phi_m$. Then, by definition
 \begin{eqnarray*}
 \widetilde f_m= \sum_{i=1} ^m \inner{ \bar{f},\phi_i} \phi_i\,,&& \Pc \bar{f}= \sum_{i=1} ^\infty \inner{ \bar{f},\phi_i} \phi_i\,,
 \end{eqnarray*}
 The polynomial
\begin{eqnarray}
\label{eq:char}\widetilde p_m(\lambda)&=& \prod_{i=1} ^m \frac{\lambda_i - \lambda}{\lambda_i}
\end{eqnarray}
fulfills $\widetilde p_m(0)=1$, hence it defines a polynomial $\widetilde q_m$ of degree $\leq m-1$ via $\widetilde p_m(\lambda)= 1 - \lambda \widetilde q_m(\lambda)\,$.  As the zeroes of $\widetilde p_m$ are the first $m$ eigenvalues of $\bar{K}$,  the polynomial $\widetilde q_m$ has the convenient property that it ''cancels out'' the first $m$ eigenfunctions, i.e.
 \begin{eqnarray*}
 \| \Pc \bar{f} - \bar{K} \widetilde q_m(\bar{K}) \bar{f}\|^2&=& \sum_{i=m+1} ^\infty \widetilde p_m(\lambda_i)^2 \inner{\bar{f}, \phi_i}^2
 \end{eqnarray*}
 By construction, $\widetilde{p}_m(\lambda_i)^2 \leq1$ for $i>m$, and hence
\begin{displaymath}
\| \Pc \bar{f} - \bar{K} \widetilde q_m(\bar{K}) \bar{f}\|^2 \leq \sum_{i=m+1} ^\infty  \inner{\bar{f}, \phi_i}^2 = \| \Pc \bar{f} - \widetilde f_m\|^2\,.
\end{displaymath}
As the principal components approximations $\widetilde f_m$ converge to $\Pc \bar{f}$,  this concludes the proof.
\end{proof}

As the rate of convergence of the population version is at least as good as the rate of the
principal components approximations, this theorem shows in particular
that the conjugate gradient method is less biased than Principal
Components Analysis. This fact is known for linear PLS in the
empirical case \citep{Jong9301,Phatak0201}. By the same token, KPLS is less biased than $\ell_2$-Boosting, as the latter corresponds to the
fixed polynomial $q_m(t) = \sum_{i=0}^{m-1}(1-t)^i$\,, . However, empirical
findings suggest that for KPLS, the decrease in bias is balanced by an
increased complexity in terms of degrees of freedom
\citep{Kraemer0701}. The goal of the next section is to
introduce a suitable control of this complexity.


\section{CONSISTENT STOPPING RULES}
\subsection{ERROR MONITORING}
We control the error between the
population case and the empirical case by iteratively monitoring
upper bounds on this error. Since this bound only involves known empirical
quantities, we design a stopping criterion based on the
bound. This then leads to a globally consistent procedure. The key ingredient of the stopping rule is to bound the differences for $u, x,d$ (defined in algorithm \ref{algo:cg1}) if we replace the empirical quantities $S$ and $T^*\y$ by their population versions. Note that  algorithm \ref{algo:cg1} involves products and quotients of the perturbed quantities. The error control based on these expressions can hence be represented in terms of the following three functions.

\begin{definition}
For any positive reals $x > \delta_x \geq 0$ define
\begin{eqnarray*}
\zeta(x,\delta_x)&=&\frac{\delta_x}{x(x-\delta_x)}
\end{eqnarray*}
 and
for any positive reals $(x,y,\delta_x,\delta_y)$ define
\begin{eqnarray*}
\xi(x,y,\delta_x,\delta_y) &= &x\delta_y + y\delta_x +
\delta_x\delta_y\,;\\
\xi'(x,y,\delta_x,\delta_y) & = & x\delta_y + y\delta_x\,.
\end{eqnarray*}
\end{definition}
The usefulness of these definitions is justified by the following
standard lemma for bounding the approximation error of inverse and
products, based only on the knowledge of the approximant:
\begin{lemma}
\label{le:control}
Let $\alpha,\bar{\alpha}$ be two invertible elements of a Banach
algebra  $\Bc$\,, with $\norm{\alpha-\bar{\alpha}}\leq \delta$\,
and $\norm{{\alpha}^{-1}}^{-1} > \delta$\,. Then
\[
\norm{\alpha^{-1} - \bar{\alpha}^{-1}} \leq \zeta(\norm{{\alpha}^{-1}}^{-1},\delta)\,.
\]
Let $\Bc_1,\Bc_2$\, be two Banach spaces  and
assume an assocative product operation exists from $\Bc_1\times\Bc_2$
to a Banach space $\Bc_3$\,, satisfying for any $(x_1,x_2)\in
\Bc_1\times\Bc_2$ the product compatibility condition $\norm{x_1 x_2}\leq \norm{x_1}\norm{x_2}$\,.
Let $\alpha,\bar{\alpha}$ in $\Bc_1$ and $\beta,\bar{\beta} \in
\Bc_2$\,, such that $\norm{\alpha-\bar{\alpha}}\leq \delta_\alpha$ and
$\|{\beta-\bar{\beta}}\|\leq \delta_\beta$\,. Then
\[
\|{\alpha\beta - \bar{\alpha}\bar{\beta}}\| \leq
\xi(\norm{{\alpha}},\|{{\beta}}\|,\delta_\alpha,\delta_\beta)\,.
\]
In the same situation as above, if it is known that
$\norm{\bar{\alpha}}\leq C$\,, then
\[
\|{\alpha\beta - \bar{\alpha}\bar{\beta}}\| \leq
\xi'(C,\|{{\beta}}\|,\delta_\alpha,\delta_\beta)\,.
\]
\end{lemma}
Furthermore, we can  bound the deviation of  the 'starting values' $S$ and $T^*\y$: 
\begin{lemma}
\label{lem:epsn}
Set $\eps_n = 4 \sqrt{(\log n)/n}\,$. If the kernel is bounded {\bf (B)}, with probability at least $1-n^{-2}$,
\begin{eqnarray*}
\| T^* \y - \bar{T} \bar{f}\|\leq  \eps_n& \text{ and } \norm{S - \bar{S}} \leq \eps_n\,.
\end{eqnarray*}
\end{lemma}
The second bound  is well-known, see
e.g. \cite{ShaCri03,ZwaBla05}\,, and the first one is based on the same argument.

The error monitoring updates corresponding to algorithm \ref{algo:cg1}
are displayed in algorithm \ref{algo:error}. Note that the error monitoring initialization and update only
depend on observable quantities.

\begin{algorithm}
\caption{Error Control for Algorithm \ref{algo:cg1}}
\begin{algorithmic}
\STATE{Initialize: $\delta^{g}_0 = 0; \eps_n = 4\sqrt{(\log n)/n}; \delta^{u}_0 =
\eps_n$;\\$ \delta^{d}_0 = \eps_n$ }
\STATE{Initialize: $\eps_{0,4} = \xi(\norm{u_0},\norm{u_0},\delta^{u}_0,\delta^{u}_0)$}
\FOR{$m=0,\ldots,(m_{\max} -1)$}
\STATE{$\eps_{m,1} = \xi'(\norm{d_m}, 1, \delta^{d}_m, \eps_n)$}
\STATE{$\eps_{m,2} = \xi(\norm{d_m}, \norm{d_m}, \delta^{d}_m,
  \eps_{m,1})$}
\STATE{$\eps_{m,3} = \zeta(\langle{d_m,S d_m}\rangle, \eps_{m,2})$ (if
defined, else exit)}
\STATE{$\delta^{\alpha}_m =
\xi(\norm{u_m}^2,\langle{d_m,Sd_m}\rangle^{-1},\eps_{m,4},\eps_{m,3})$}
\STATE{$\delta^{g}_{m+1} = \delta^{g}_m + \xi(\alpha_m,\norm{d_m},
\delta^{\alpha}_m,\delta^{d}_m)$}
\STATE{$\delta^{u}_{m+1} = \delta^{u}_m + \xi(\alpha_m,\norm{d_m},
\delta^{\alpha}_m, \eps_{m,1})$}
\STATE{$\eps_{m,5} = \zeta(\norm{u_m}^2,\eps_{m,4})$ (if defined, else
exit)}
\STATE{$\eps_{m+1,4} =
  \xi(\norm{u_{m+1}},\norm{u_{m+1}},\delta^{u}_{m+1},\delta^{u}_{m+1})$}
\STATE{$\delta^{\beta}_m =
\xi(\norm{u_{m+1}}^2,\norm{u_{m}^{-1}}^2,\eps_{m+1,4},\eps_{m,5})$}
\STATE{$\delta^{d}_{m+1} = \delta^{d}_m +
\xi(\beta_m,\norm{d_m},\delta^{\beta}_m,\delta^{d}_m)$}
\ENDFOR
\end{algorithmic}
\label{algo:error}
\end{algorithm}
\begin{definition}[First stopping rule]
Fix $0<\gamma<\frac{1}{2}$ and run the KPLS algorithm \ref{algo:cg1}
along with the error monitoring algorithm \ref{algo:error}. Let $m_{(n)}+1$ denote the
first time where either the procedure exits, or $\delta^g_{m_{(n)}+1}
> n^{-\gamma}$\,. Here, the subscript $(n)$ indicates
 that the step is data-dependent. Output the
estimate at the previous step, that is, the estimate $f^{(n)} =T g_{m_{(n)}}$.
\end{definition}
The next theorem states that this stopping rule leads to a
universally consistent learning rule for bounded regression.

\begin{theorem}
\label{thm:cons1}
Assume that the kernel is bounded {\bf (B)} and universal {\bf (U)}.
Denote by $f^{(n)}$ the output of
the KPLS run on $n$ independent observations while using the above stopping rule. Then almost surely
\[
\lim_{n\rightarrow \infty} \norm{\bar{f}-f^{(n)}}_{\mathcal{L}_2(P_X)} =0\,.
\]
\end{theorem}
\begin{proof}
By construction $f^{(n)} = T g^{(n)}_{m_{(n)}}$\,.
We have
\[
\norm{\bar{f}-f^{(n)}} \leq \norm{\bar{f} - \bar{f}_{m_{(n)}}} +
\norm{\bar{f}_{m_{(n)}}-Tg^{(n)}_{m_{(n)}}}\,.\]
We proceed in two steps. First, we establish that the second term is
bounded by $n^{-\gamma}$. Second, we prove that the random
variable $m_{(n)}\rightarrow \infty$ almost surely for $n\rightarrow \infty$, which ensures
that the first term goes to zero (using Theorem \ref{th:popcase}).

\emph{First Step:} We  have
\begin{eqnarray*}
\norm{\bar{f}_{m_{(n)}}-Tg^{(n)}_{m_{(n)}}}_{\mathcal{L}_2(P_X)}&=& \norm{T(\bar{g}_{m} -  g_{m})}_{\mathcal{L}_2(P_X)} \\
&\leq &\norm{\bar{g}_{m} -  g_{m}}_\infty  \\
&\leq &\norm{\bar{g}_{m} -  g_{m}}_{\Hc_k}\,.
\end{eqnarray*}
(We drop the superscript $(n)$ for $m$ to lighten notation.)
Now, we prove that the construction of the error monitoring
iterates ensures that $\norm{\bar{g}_{m} -  g_{m}} \leq
\delta^{g}_m$\, for any $m$ before stopping, with probability at
least $1- 2n^{-2}$. For $m=0$, this follows immediately from Lemma \ref{lem:epsn}.
For a general $m$, it is then straightforward to show that $\eps_{m,1}$ controls the error for $S d_m$
(using
$\norm{S}\leq 1$); $\eps_{m,2}$ controls the error for $\langle
\bar{d}_k, S \bar{d}_k \rangle$ (the Cauchy-Schwartz inequality
ensuring the compatibility condition of norm and product in Lemma
\ref{le:control}); $\eps_{m,3}$ controls the error for the
inverse of the latter;
$\delta^{\alpha}_m,\delta^{g}_m,\delta^{u}_m,\delta^{\beta}_m,\delta^{d}_m,$
control the errors for the respective superscript quantities;
$\eps_{m,4}$ controls the error for $\norm{\bar{u}_m}^2$ and
$\eps_{m,5}$ for the inverse of the latter. In particular, with probability at least $1-2n^{-2}$ we have $\norm{g_m - \bar{g}_m} \leq \delta^{g}_m\,,$
 so that
$\norm{g_{m_{(n)}} - \bar{g}_{m_{(n)}}} \leq n^{-\gamma}$\, by definition
of the stopping rule. Since the probabilities that this inequalities
are violated are summable, by
the Borel-Cantelli lemma, almost surely this inequality is true from a
certain rank $n_0$ on.

\emph{Second Step: } This is in essence a minutely detailed continuity argument. Due to space limitations, we only sketch the proof. We consider a fixed
iteration $m$\, and prove that this iteration is reached without
exiting and that $\delta^{g,(n)}_m < C(m) \eps_n$ almost surely from a
certain rank $n_0$ on (here the superscript $(n)$ once again recalls
the dependence on the number of data). The constant $C(m)$ is
deterministic but can depend on the generating distribution. This obviously ensures
$m_{(n)} \rightarrow \infty$ almost surely. We prove by induction that this property is true for all
error controlling quantities $\eps_{*,m}$
and $\delta^*_m$ appearing in the error monitoring
iterates. Obviously, from the initialization this is true for
$\delta^{g}_0,\delta^{u}_0,\delta^{d}_0$ and $\eps_n$\,. In a nutshell, the induction step then follows from the fact that the error monitoring functions $\zeta,\xi,\xi'$ are locally Lipschitz.
\end{proof}

\subsection{EMPIRICAL COMPLEXITY}
We now propose an alternate stopping rule directly based on the
characterizing property \eqref{eq:cg_crit} of KPLS/CG.
This approach leads to   a
more explicit stopping rule.
Let us define the operator $R_m: \mbr^m \rightarrow \Hc $:
\begin{eqnarray}
\label{eq:Rm}
v=(v_1,\ldots,v_m) \mapsto R_m v = \sum_{i=1}^m v_i S^{i-1} T^* \y\,.
\end{eqnarray}
Then, we can rewrite the $m$-th iterate of KPLS/CG as $g_m = R_m w_m$\,,
and \eqref{eq:cg_crit} becomes
\[
w_m =  \text{arg}\min_{w \in\mbr^m} \| \y - T R_m w\|^2\,.
\]
From this, it can be deduced by standard arguments that
\begin{eqnarray}
\label{eq:xm_closed}
g_m= R_m M_m^{-1} R_m^*  T^* \y&,\, & M_m = R_m ^* S R_m\,.
\end{eqnarray}
The random $m\times m$ matrix $M_m$ has $(i,j)$ entry
\[
\left(M_m\right)_{ij} = \y^\top T S^{i-1} S S^{j-1} T^* \y = \y^\top
K^{(i+j)} \y\,,
\]
where $K=TT^*$ can be identified with the kernel Gram matrix, i.e.~$K_{k\ell}=k(x_k,x_\ell)$\,.
Similarly, denote by $M'_m$ the matrix with entry $(i,j)$ equal to
$\y^\top K^{(i+j-1)} \y$\,.
\begin{definition}[Second stopping rule]
Fix $\frac{1}{2} > \nu >0 $  and
let $m' _{(n)}+1$ denote the first time where $m \paren{ \max(\norm{M'_m},m^{-1}) \norm{M_m^{-1}}}^2\geq
n^\nu$\,. (If $M_m$ is singular, we set $\norm{M_m^{-1}} = \infty$\,.) Output the KPLS estimate at step $m' _{(n)}$\,.
\end{definition}
 Note that the stopping criterion only depends on empirical quantities. Furthermore, we can interpret the stopping criterion as an empirical complexity control of the CG algorithm at step $m$. Essentially, it  is the dimensionality (or iteration number) $m$ times the square of the ``pseudo-condition number''
$\norm{M'_m}\norm{M_m^{-1}}$\,.

\begin{theorem}
\label{altconst}
Assume that the kernel is bounded {\bf (B)} and universal {\bf (U)}.
Denote by $f^{(n)}$ the output of
the KPLS algorithm run on $n$ independent observations while using the
above stopping rule. Then almost surely
\[
\lim_{n\rightarrow \infty} \norm{\bar{f} -f^{(n)}}_{\mathcal{L}_2(P_X)} =0\,.
\]
\end{theorem}
\begin{proof}
We prove this result by exploiting the explicit formula \eqref{eq:xm_closed} for KPLS. Note that the formula is also true for the population version $\bar{x}_m$ if we replace all empirical quantities by their population counterparts. Hence, we need  to control the deviation of the different terms
appearing in the above formula (\ref{eq:xm_closed}). This boils down
to perturbation analysis and is closely related to techniques
employed by \cite{Chun0901} in a finite dimensional context. The following lemma summarizes the deviation control (the proof can be found in the appendix).
\begin{lemma}
\label{alterrorctrl}
Put $\eps_n = 4 \sqrt{(\log n)/n}$\,.
There is an absolute numerical constant $c$ such that, with
probability at least $1-n^{-2}$\,, whenever
\begin{eqnarray*}
c m \eps_n \paren{ \max(\norm{M'_m},m^{-1}) \norm{M_m^{-1}}}^2 &\leq &1\,,
\end{eqnarray*}
then
\[
\norm{g_m - \bar{g}_m} \leq c m \eps_n \paren{ \max\paren{\norm{M'_m},m^{-1}} \norm{M_m^{-1}}}^2\,.
\]
\end{lemma}

Now, let $c$ be the universal constant appearing in Lemma
\ref{alterrorctrl}. At the stopping step $m = m' _{(n)}$, by
construction,
\[
c m \eps_n  (\max(\norm{M'_m},m^{-1}) \norm{M_m^{-1}})^2 \leq 4 c
\frac{n^{\nu-\frac{1}{2}}}{\sqrt{\log n}} \,.
\]
Choosing some positive $\gamma < \frac{1}{2} - \nu$\,, for $n$ large enough
the right hand side is bounded by $n^{-\gamma}$\,. By Lemma
\ref{alterrorctrl} and a reasoning similar to that of the proof
of Theorem \ref{thm:cons1}, this implies the bound on the estimation error
with respect to the population version (valid with probability at least $1-n^{-2}$):
\[
\norm{\bar{f}_{m'_{(n)}} - f_{m' _{(n)}}} \leq n^{-\gamma}\,.
\]
By the Borel-Cantelli lemma, this inequality is therefore almost
surely satisfied for big enough $n$\,.

To conclude, we establish that $m' _{(n)} \rightarrow \infty$
almost surely as $n\rightarrow \infty$\,. It suffices to show
that for any fixed $m$\,, for $n$ larger than a certain $n_0(m)$ on,  $m (\max(\norm{M'_m},m^{-1}) \norm{M_m^{-1}})^2\leq n^\nu$, implying that almost surely $m'_{(n)} \geq m$\, for $n$ large enough. For this we simply show that the LHS of this inequality converges a.s. to a fixed number. From \eqref{eq:convm} in the proof of Lemma \ref{alterrorctrl}, and the straightforward inequality
$\norm{M'}\leq m$\,, we see that for a fixed iteration $m$\,,
$\norm{M_m-\bar{M}_m} \rightarrow 0$\, almost surely as $n\rightarrow infty$. The matrix $\bar{M}_m$ is non-singular
as we assume that the population version of the algorithm does
not exit. This implies that $M_m$ is almost surely non-singular for $n$ big enough
with $\norm{M_m^{-1} - \bar{M}_m^{-1}} \rightarrow 0$\,, and
therefore almost surely $\norm{M_m^{-1}}$ converges to
$\norm{\bar{M}_m^{-1}}$\,, a fixed number.
\end{proof}

\section{CONCLUSION}
In this paper, we proposed two stopping rules for the number $m$ of KPLS iterations that lead to universally consistent estimates. Both are based on the facts that  (a) population KPLS is defined in terms of the covariance operator $\bar{S}$ and the cross-covariance $\bar{T}^* \bar{f}$, and that (b) we can control the difference between population KPLS and empirical KPLS solutions  based on the discrepancy of these covariances to their empirical counterparts $S$ and $T^* \y$. While the first stopping rule monitors the estimation error  by following the iterative KPLS algorithm \ref{algo:cg1}, the second stopping rule uses a closed-form representation and is expressed in terms of a pseudo condition number. Both rules do not require any prior knowledge on the target function and can be computed from the data.

Our approach makes heavy use of the equivalence of PLS to the conjugate gradient algorithm applied to the normal equations in combination with early stopping. This framework also connects KPLS to statistical inverse problems. In this context, KPLS stands out of previously well-studied classes of methods. In particular, as KPLS is not linear in $\y$, it contrasts the class of \emph{linear} spectral methods for statistical inverse problems (e.g. \citealp{Bissantz0701,LogRosetal08}). This class considers estimates of the form
\begin{eqnarray*}
g_{\lambda}&=& F_{\lambda}(S) T^* \y
\end{eqnarray*}
in order to regularize the ill-posed problem \eqref{eq:normal_emp}. Here, $F_{\lambda}$ is a fixed family of ''filter functions''. Examples include Ridge Regression (also known as Thikonov regularization), Principal Components Regression (also known as spectral cut-off), and $\ell_2$-Boosting (which corresponds to Landweber iteration). In this paper, we explained that for KPLS with $m$ components, the filter function $F_{\lambda}$ is a polynomial of degree $\leq m-1$, but that -- unlike the class of linear spectral methods -- this polynomial strongly depends on the response. For this reason, general results for linear spectral methods do not directly apply to KPLS, and additional techniques are needed.

For linear spectral methods, optimal convergence rates of the resulting estimators have been established in the recent years
\citep{CapDeV07,BauPerRos07,Cap06}. Obviously, beyond consistency the focus of future theoretical effort on KPLS will be to establish that this algorithm can attain these optimal rates as well.



\bibliographystyle{apalike}

\appendix
\section{PROOF OF LEMMA \ref{alterrorctrl}}
The letter $c$ denotes an absolute numerical constant whose value
can possibly be different form line to line.

We define $\bar{R}_m$ as the
population analogue of $R_m$\,, by replacing $S$ and $T^*\y$ in \eqref{eq:Rm} by $\bar{S}$ and $\bar{T}^*\bar{f}$.
The population version of KPLS is then
\[
\bar{x}_m =  \bar{R}_m \bar{M}_m^{-1} \bar{R}_m^*  \bar{T}^* \bar{f}
\]
where $\bar{M}_m = \bar{R}_m ^* \bar{S} \bar{R}_m $\,.
Since the index $m$ of the iteration
is now fixed, we omit the subscript $m$ in $R$ and $M$. Set $\Delta = \max\paren{\norm{M'},m^{-1}}$\,.
We have $M'=RR^*$\,, so that $\norm{M'} = \norm{R}^2$\,. Observe that
\[
M'-M = R_m(Id-S)R_m^*\,,
\]
is a positive-semidefinite matrix since $\norm{S}\leq 1$; hence $\norm{M'} \geq \norm{M} \geq \norm{M^{-1}}^{-1}$,
and the assumption that $C \eps_n m \paren{ \Delta \norm{M^{-1}}}^2 \leq 1$ with $C\geq 1$ implies in particular
that $m \eps_n \leq 1$\,.

We first control the difference $\rho=\norm{R - \bar{R}}$\,.
\begin{eqnarray*}
\rho&=& \sup_{v:\norm{v}=1} \norm{ \sum_{i=1}^m v_i \paren{S^{i-1} T^* \y -
      \bar{S}^{i-1} \bar{T}^* \bar{f}}}\\
      &\leq&\sup_{v:\norm{v}=1} \paren{\norm{ \sum_{i=1}^m v_i
      \paren{S^{i-1}  - \bar{S}^{i-1}} T^* \y} }\\
&&+ \sup_{v:\norm{v}=1} \norm{ \sum_{i-1}^m v_i \bar{S}^{i-1} \paren{ T^* \y -
      \bar{T^*}\bar{f}}}\,.
\end{eqnarray*}
Using lemma \ref{lem:epsn}, the second term on the right-hand side can be bounded by  $\sum_{i=1}^m \abs{v_i} \eps_n  \leq \norm{v} \sqrt{m} \eps_n\,$.
For the first term, one can rewrite
\begin{multline*}
\sup_{v:\norm{v}=1} \norm{ \sum_{i=1}^m v_i
      \paren{S^{i-1}  - \bar{S}^{i-1}} T^* \y}  \\
\begin{aligned}
& = \sup_{v:\norm{v}=1} \norm{ \sum_{i,j=1}^{m-1}
      {\bf 1}\{j\leq i\} v_i \bar{S}^{j-1}(S-\bar{S}) S^{i-j} T^* \y}  \\
& = \sup_{v:\norm{v}=1} \norm{ \sum_{j=1}^{m-1}
     \bar{S}^{j-1}(S-\bar{S}) \sum_{i=j}^{m-1} v_i S^{i-j} T^* \y}  \\
& \leq \sum_{j=1}^{m-1} \eps_n   \sup_{v:\norm{v}=1}
     \norm{\sum_{i=1}^{m-j} v_i S^{i-1} T^* \y}  \leq m \eps_n \norm{R}\,.
\end{aligned}
\end{multline*}
This implies
\[
\norm{R -\bar{R}} \leq \sqrt{m} \eps_n + m \eps_n \norm{R}
\leq 2 m \eps_n \Delta^{\frac{1}{2}}\,. 
\]
We now use repeatedly Lemma~\ref{le:control} to derive the final
estimate from the above one using product and inverse operations.
We omit some tedious details in the computations below.

Recalling $M=R ^* S  R$ and using $m \eps_n \leq 1$\,, we deduce
\begin{equation}
\label{eq:convm}
\norm{M - \bar{M}} \leq  c \paren{m \eps_n \Delta + m^2 \eps_n^2 \Delta } \leq  c m \eps_n \Delta\,.
\end{equation}
We then have
\begin{align*}
\norm{M^{-1} - \bar{M}^{-1} }  \leq \zeta(\norm{M^{-1}}^{-1}, c m \eps_n \Delta) \leq c m \eps_n \Delta \norm{M^{-1}}^2\,.
\end{align*}
Here, we assume that $c m \eps_n \Delta \leq \norm{M^{-1}}^{-1}/2$\,,
which is implied by the assumption $C\eps_n m \Delta^2 \norm{M^{-1}}^2 \leq 1$ if
$C$ is chosen big enough. Finally, we get
\[
\norm{R M^{-1} R^* - \bar{R} \bar{M}^{-1} \bar{R}^* }
\leq c m \eps_n \Delta^2 \norm{M^{-1}}^2\,.
\]

\end{document}